\documentclass[nonacm, sigconf, twocolumn]{acmart}
\renewcommand\footnotetextcopyrightpermission[1]{}
\setcopyright{none}

\usepackage{booktabs}
\usepackage{graphicx}
\usepackage{algorithm}
\usepackage[noend]{algpseudocode}
\usepackage{multirow}
\usepackage{graphicx}
\usepackage{amsmath}

\usepackage{url}

\usepackage{breakurl}

\usepackage{float}

\usepackage{xspace}
\usepackage{subfig}
\usepackage{xcolor}
\usepackage{tabularx,colortbl}
\usepackage[inline]{enumitem}

\usepackage{hyperref}
\hypersetup{
  colorlinks=true,      
  linkcolor=blue,       
  citecolor=magenta,    
  filecolor=cyan,       
  urlcolor=red          
}

\usepackage[compact,small]{titlesec}
\usepackage[font={small}, skip=1pt]{caption}

\usepackage{algorithm, algorithmicx,algpseudocode}
\algrenewcommand\algorithmicindent{0.5em}%

\newenvironment{denseitemize}{
\begin{itemize}[topsep=2pt, partopsep=0pt, leftmargin=1em]
  \setlength{\itemsep}{2pt}
  \setlength{\parskip}{0pt}
  \setlength{\parsep}{0pt}
}{\end{itemize}}

\def\ie{{i.e.\xspace}}
\def\eg{{e.g.\xspace}}

\def\etc{etc.\xspace}

\def\infiniswap{Infiniswap\xspace}
\def\leap{Leap\xspace}
\def\hydra{Hydra\xspace}
\def\memtrade{Memtrade\xspace}
\def\tpp{TPP\xspace}
\def\justitia{Justitia\xspace}
\def\chameleon{Chameleon\xspace}

\def\cxlmem{CXL-Memory\xspace}

\begin{document}
\sloppy
\date{}

\title{\huge Memory Disaggregation: Advances and Open Challenges}

\author{\large Hasan Al Maruf, Mosharaf Chowdhury} \affiliation{\institution{SymbioticLab, University of Michigan}}

\begin{abstract}
Compute and memory are tightly coupled within each server in traditional datacenters. 
Large-scale datacenter operators have identified this coupling as a root cause behind fleet-wide resource underutilization and increasing Total Cost of Ownership (TCO). 
With the advent of ultra-fast networks and cache-coherent interfaces, \emph{memory disaggregation} has emerged as a potential solution, whereby applications can leverage available memory even outside server boundaries.

This paper summarizes the growing research landscape of memory disaggregation from a software perspective and introduces the challenges toward making it practical under current and future hardware trends. 
We also reflect on our seven-year journey in the SymbioticLab to build a comprehensive disaggregated memory system over ultra-fast networks. 
We conclude with some open challenges toward building next-generation memory disaggregation systems leveraging emerging cache-coherent interconnects.
\end{abstract}

\maketitle
\thispagestyle{empty}
\section{Introduction}
\label{sec:intro}

Modern datacenter applications -- low-latency online services, big data analytics, and AI/ML workloads alike -- are often memory-intensive. 
As the number of users increases and we collect more data in cloud datacenters, the overall memory demand of these applications continue to rise. 
%
%
Despite their performance benefits, memory-intensive applications experience \emph{disproportionate} performance loss whenever their working sets do not completely fit in the available memory.
For instance, our measurements across a range of memory-intensive applications show that if half their working sets do not fit in memory, performance can drop by $8\times$ to $25\times$ \cite{infiniswap}. 

Application developers often sidestep such disasters by over-allocating memory, but pervasive over-allocation inevitably leads to datacenter-scale memory underutilization.
Indeed, memory utilization at many hyperscalers hovers around 40\%--60\% \cite{google-trace-analysis, alibaba-cluster-trace, infiniswap, googledisagg}.
Service providers running on public clouds, such as Snowflake, report 70\%--80\% underutilized memory on average \cite{snowset}.
Since DRAM is a significant driver of infrastructure cost and power consumption \cite{tpp}, excessive underutilization leads to high TCO.

At the same time, increasing the effective memory capacity and bandwidth of each server to accommodate ever-larger working sets is challenging as well.  
In fact, memory bandwidth is a bigger bottleneck than memory capacity today as the former increases at a slower rate.
For example, to increase memory bandwidth by $3.6\times$ in their datacenters, Meta had to increase capacity by $16\times$ \cite{tpp}.
To provide sufficient memory capacity and/or bandwidth, computing and networking resources become stranded in traditional server platforms, which eventually causes fleet-wide resource underutilization and increases TCO. 

\emph{Memory disaggregation} addresses memory-related rightsizing problems at both software and hardware levels. 
Applications are able to allocate memory as they need without being constrained by server boundaries. 
Servers are not forced to add more computing and networking resources when they only need additional memory capacity or bandwidth.
By exposing all unused memory across all the servers as a memory pool to all memory-intensive applications, memory disaggregation can improve both application-level performance and overall memory utilization. 
Multiple hardware vendors and hyperscalers have projected \cite{intel-yahoo-rasc-savings, intel-tencent-rasc-savings, pond-azure, tpp} up to 25\% TCO savings without affecting application performance via (rack-scale) memory disaggregation.

While the idea of leveraging remote machines' memory is decades old \cite{transparent-remote-paging-vm, cashmere-vlm, zahorjan-remote-paging, hpnbd, markatos-remote, nswap}, only during the past few years, the latency and bandwidth gaps between memory and communication technologies have come close enough to make it practical. 
The first disaggregated memory\footnote{Remote memory and far memory are often used interchangeably with the term disaggregated memory.} solutions (Infiniswap \cite{infiniswap} and the rest) leveraged RDMA over InfiniBand or Ethernet, but they are an order-of-magnitude slower than local memory. 
To bridge this performance gap and to address practical issues like performance isolation, resilience, scalability, \etc, we have built a comprehensive set of software solutions.
More recently, with the rise of cache-coherent Compute Express Link (CXL) \cite{cxl} interconnects and hardware protocols, the gap is decreasing even more.
We are at the cusp of taking a leap toward next-generation software-hardware co-designed disaggregated memory systems. 

This short paper is equal parts a quick tutorial, a retrospective on the Infiniswap project summarizing seven years' worth of research, and a non-exhaustive list of future predictions based on what we have learned so far. 

\section{Memory Disaggregation}
\label{sec:overview}
Simply put, memory disaggregation exposes memory capacity available in remote locations as a pool of memory and shares it across multiple servers over the network. 
It decouples the available compute and memory resources, enabling independent resource allocation in the cluster. 
A server's local and remote memory together constitute its total physical memory. 
An application's locality of memory reference allows the server to exploit its fast local memory to maintain high performance, while remote memory provides expanded capacity with an increased access latency that is still orders-of-magnitude faster than accessing persistent storage (\eg, HDD, SSD).
The OS and/or application runtime provides the necessary abstractions to expose all the available memory in the cluster, hiding the complexity of setting up and accessing remote memory (\eg, connection setup, memory access semantics, network packet scheduling, \etc) while providing resilience, isolation, security, \etc guarantees. 

\subsection{Architectures}

\begin{figure}[!t]
	\centering
	\subfloat[][\textbf{Physically Disaggregated}]{%
		\label{fig:physical}%
		\includegraphics[width=0.6\columnwidth]{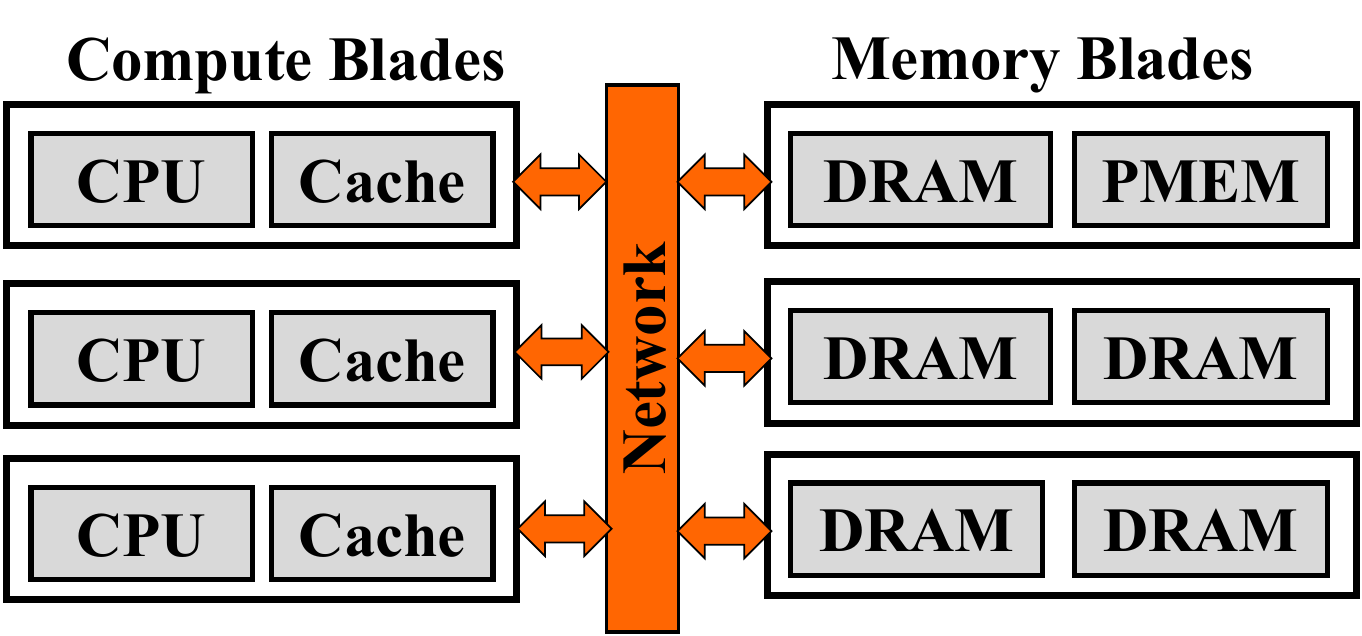}%
	}
  
  \subfloat[][\textbf{Logically Disaggregated}]{%
		\label{fig:logical}%
		\includegraphics[width=0.6\columnwidth]{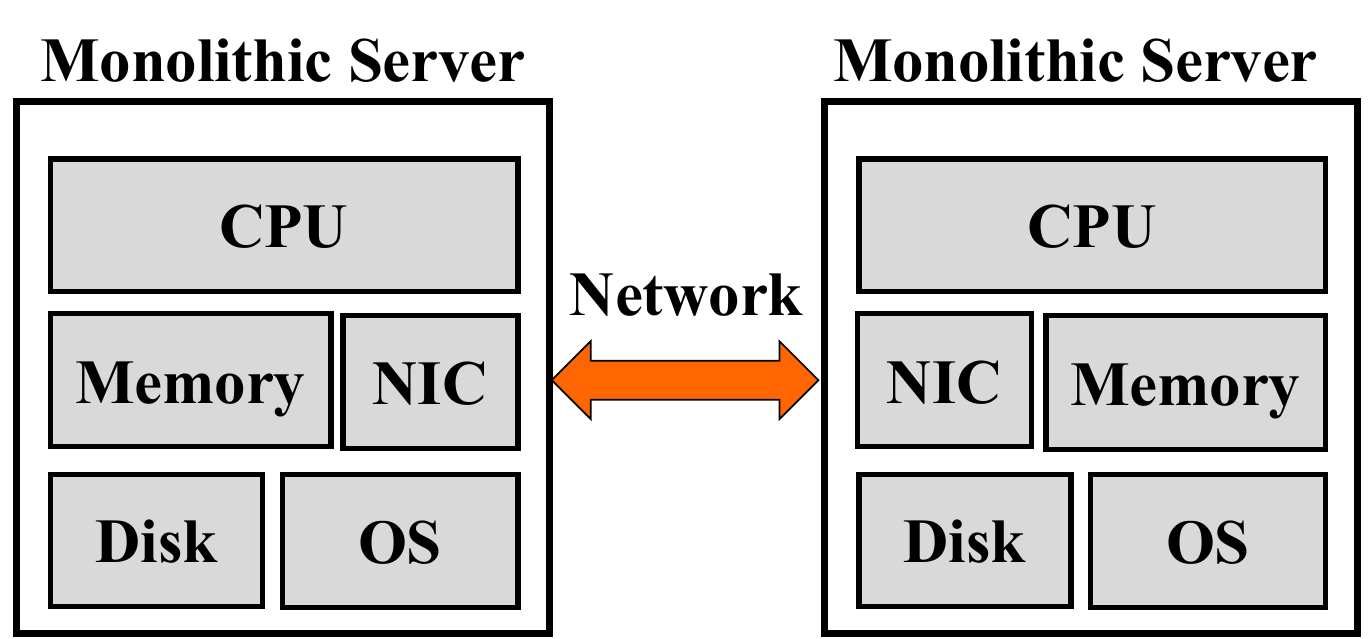}%
	}	
	\caption{Physical vs. logical disaggregation architectures.} 
  \label{fig:disaggregation_architecture}
\end{figure}

Memory disaggregation systems have two primary cluster memory architectures. 

\paragraph{Physical Disaggregation.}
In a physically-disaggregated architecture, compute and memory nodes are detached from each other where a cluster of compute blades are connected to one or more memory blades through network (\eg, PCIe bridge) \cite{bladedisagg} (Figure~\ref{fig:physical}).
A memory node can be a traditional monolithic server with low compute resource and large memory capacity, or it can be network-attached DRAM.
For better performance, the compute nodes are usually equipped with a small amount of memory for caching purposes.

\paragraph{Logical Disaggregation.}
In a logically-disaggregated architecture, traditional monolithic servers hosting both compute and memory resources are connected to each other through the network (\eg, Infiniband, RoCEv2) (Figure~\ref{fig:logical}).
This is a popular approach for building a disaggregated memory system because one does not need to change existing hardware architecture; simply incorporating appropriate software to provide a remote memory interface is sufficient.
In such a setup, usually, each of the monolithic servers has their own OS.
In some cases, the OS itself can be disaggregated across multiple hosts \cite{legoos}.
Memory local to a host is usually prioritized for running local jobs. 
Unutilized memory on remote machines can be pooled and exposed to the cluster as remote \cite{infiniswap, leap, hydra, memtrade, legoos, AIFM, kona}. 

\paragraph{Hybrid Approach.}
Cache-coherent interconnects like CXL provides the opportunity to build a composable heterogeneous memory systems that combine logical and physical disaggregation approaches. 
Multiple monolithic servers, compute devices, memory nodes, or network specialized devices can be connected through fabric or switches where software stacks can provide the cache-line granular or traditional virtual memory-based disaggregated memory abstraction.


\begin{table}[]
  \caption{Selected memory disaggregation proposals.}
  \label{tab:related-work}
  \resizebox{\columnwidth}{!}{%
\begin{tabular}{|c|c|c|c|c|}
\hline
Abstraction &
  System &
  \begin{tabular}[c]{@{}c@{}}Hardware \\ Transparent\end{tabular} &
  \begin{tabular}[c]{@{}c@{}}OS \\ Transparent\end{tabular} &
  \begin{tabular}[c]{@{}c@{}}Application \\ Transparent\end{tabular} \\ \hline
\multirow{9}{*}{\begin{tabular}[c]{@{}c@{}}Virtual \\ Memory \\ Management \\ (VMM)\end{tabular}} &
  Global Memory \cite{globalmemory95} &
  Yes &
  No &
  Yes \\ \cline{2-5} 
                                                                                & Memory Blade \cite{bladedisagg}   & No  & No  & Yes \\ \cline{2-5} 
                                                                                & Infiniswap \cite{infiniswap}     & Yes & Yes & Yes \\ \cline{2-5} 
                                                                                & Leap \cite{leap}           & Yes & No  & Yes \\ \cline{2-5} 
                                                                                & LegoOS \cite{legoos}         & Yes & No  & Yes \\ \cline{2-5} 
                                                                                & zSwap \cite{googledisagg}          & Yes & No  & Yes \\ \cline{2-5} 
                                                                                & Kona \cite{kona}           & Yes & No  & Yes \\ \cline{2-5} 
                                                                                & Fastswap \cite{fastswap}       & Yes & No  & Yes \\ \cline{2-5} 
                                                                                & Hydra \cite{hydra}          & Yes & Yes & Yes \\ \hline
\multirow{2}{*}{\begin{tabular}[c]{@{}c@{}}Virtual File \\ System (VFS)\end{tabular}} &
  Memory Pager \cite{markatos-remote} &
  Yes &
  Yes &
  No \\ \cline{2-5} 
& Remote Regions \cite{remote-regions} & Yes & Yes & No  \\ \hline
\multirow{3}{*}{\begin{tabular}[c]{@{}c@{}}Custom \\ API\end{tabular}}          
& FaRM \cite{farm}           & Yes & Yes & No  \\ \cline{2-5} 
 & FaSST \cite{fasst}           & Yes & Yes & No  \\ \cline{2-5} 
& Memtrade \cite{memtrade}       & Yes & Yes & No  \\ \hline
\multirow{2}{*}{\begin{tabular}[c]{@{}c@{}}Programming \\ Runtime\end{tabular}} & AIFM \cite{AIFM}           & Yes & Yes & No  \\ \cline{2-5} 
& Semeru \cite{semeru}         & Yes & Yes & No  \\ \hline 
\end{tabular}%
}
\end{table}

\subsection{Abstractions and Interfaces}

Interfaces to access disaggregated memory can either be transparent to the application or need minor to complete re-write of applications (Table~\ref{tab:related-work}). 
The former has broader applicability, while the latter might have better performance.

\paragraph{Application-Transparent Interface.}
Access to remote disaggregated memory without significant application rewrites typically relies on two primary mechanisms: disaggregated Virtual File System (VFS) \cite{remote-regions}, that exposes remote memory as files and disaggregated Virtual Memory Manager (VMM) for remote memory paging \cite{leap, infiniswap, hydra, legoos}. 
In both cases, data is communicated in small chunks or pages (typically, 4KB). 
In case of remote memory as files, pages go through the file system before they are written to/read from the remote memory. 
For remote memory paging and distributed OS, page faults cause the VMM to write pages to and read them from the remote memory.
Remote memory paging is more suitable for traditional applications because it does not require software or hardware modifications.

\paragraph{Non-Transparent Interface.} 
Another approach is to directly expose remote memory through custom API (KV-store, remote memory-aware library or system calls) and modify the applications incorporating these specific APIs \cite{AIFM, semeru, fasst, farm, nu, memtrade}.
All the memory (de)allocation, transactions, synchronizations, \etc operations are handled by the underlying implementations of these APIs.
Performance optimizations like caching, local-vs-remote data placement, prefetching, \etc are often handled by the application. 

\begin{figure}[!t]
  \centering
  \includegraphics[width=\columnwidth]{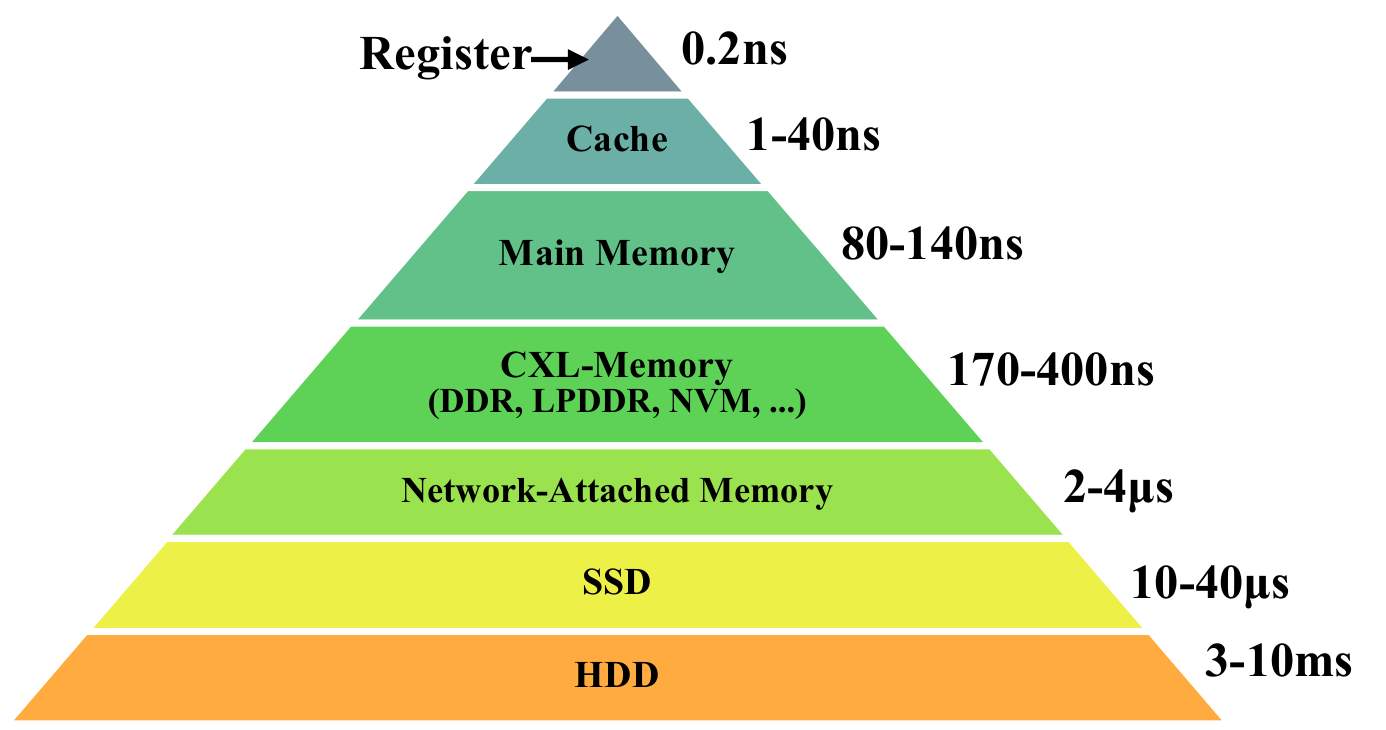}
  \caption{Latency profile of different memory technologies.}
  \label{fig:tiered-memory}
  \vspace{-2em}
\end{figure}

\subsection{Challenges in Practical Memory Disaggregation}


Simply relying on fast networks or interconnects is not sufficient to practical memory disaggregation.
A comprehensive solution must address challenges in multiple dimensions:

\begin{denseitemize}
\item \textbf{High Performance.} 
A disaggregated memory system involves the network in its remote memory path, which is at least an order-of-magnitude slower than memory channels attached to CPU and DRAM (80--140 nanoseconds vs. microseconds; see Figure~\ref{fig:tiered-memory}).
Hardware-induced remote memory latency is significant and impacts application performance \cite{infiniswap, leap, tpp}.
Depending on the abstraction, software stacks can also introduce significant overheads. 
For example, remote memory paging over existing VMM can add tens of microseconds latency for a 4KB page \cite{leap}.



\item \textbf{Performance Isolation.}
When multiple applications with different performance requirements (\eg, latency- vs. bandwidth-sensitive workloads) compete for disaggregated memory, depending on where the applications are running and where the remote memory is located, they may be contending for resources inside the server, on the NIC, and in the network on the hardware side and variety of resources in the application runtimes and OSes. 
This is further exacerbated by the presence of multiple tiers of memory with different latency-bandwidth characteristics. 


\item \textbf{Memory Heterogeneity.} 
Memory hierarchy within a server is already heterogeneous (Figure~\ref{fig:tiered-memory}).
Disaggregated memory -- both network-attached and emerging CXL memory \cite{pond-azure, directcxl, tpp} -- further increases heterogeneity in terms of latency-bandwidth characteristics. 
In such a setup, simply allocating memory to applications is not enough.
Instead, decisions like how much memory to allocate in which tier at what time is critical as well. 

\item \textbf{Resilience to Expanded Failure Domains.} 
Applications relying on remote memory become susceptible to new failure scenarios such as independent and correlated failures of remote machines, evictions from and corruptions of remote memory, and network partitions.
They also suffer from stragglers or late-arriving remote responses due to network congestion and background traffic \cite{tail-scale}.
These uncertainties can lead to catastrophic failures and service-level objective (SLO) violations.

\item \textbf{Efficiency and Scalability.}
Disaggregated memory systems are inherently distributed. 
As the number of memory servers, the total amount of disaggregated memory, and the number of applications increase, the complexity of finding unallocated remote memory in a large cluster, allocating them to applications without violating application-specific SLOs, and corresponding meta-data overhead of memory management increase as well. 
Finding efficient matching at scale is necessary for high overall utilization.

\item \textbf{Security.} 
Although security of disaggregated memory is often sidestepped within the confines of a private datacenter, it is a major challenge for memory disaggregation in public clouds.
Since data residing in remote memory may be read by entities without proper access, or corrupted from accidents or malicious behavior, the confidentiality and integrity of remote memory must be protected. 
Additional concerns include side channel and remote rowhammer attacks over the network \cite{pythia, throwhammer}, distributed coordinated attacks, lack of data confidentiality and integrity and client accountability during CPU bypass operations (\eg, when using RDMA for memory disaggregation).

\end{denseitemize} 

\section{Infiniswap: A Retrospective}
\label{sec:infinix}

To the best of our knowledge, Infiniswap is the first memory disaggregation system with a comprehensive and cohesive set of solutions for all the aforementioned challenges. 
It addresses host-level, network-level, and end-to-end aspects of practical memory disaggregation over RDMA. 
At a high level, Infiniswap provides a paging-based remote memory abstraction that can accommodate any application without changes, while providing a high-performance yet resilient, isolated, and secure data path to remote disaggregated memory.

\paragraph{Bootstrapping.} 
Our journey started in 2016, when we simply focused on building an application-transparent interface to remote memory that are distributed across many servers. 
{\infiniswap} \cite{infiniswap} transparently exposed remote disaggregated memory through paging without any modifications to applications, hardware, or OSes of individual servers.
It employed a block device with traditional I/O interface to VMM.
The block device divided its whole address space into smaller slabs and transparently mapped them across many servers' remote memory.
Infiniswap captured 4KB page faults in runtime and redirected them to remote memory using RDMA.

From the very beginning, we wanted to design a system that would scale without losing efficiency down the line.
To this end, we designed decentralized algorithms to identify free memory, to distribute memory slabs, and to evict slabs for memory reclamation.
This removed the overhead of centralized meta-data management without losing efficiency. 


\paragraph{Improving Performance.} 
\infiniswap's block layer-based paging caused high latency overhead during remote memory accesses.
This happens because Linux VMM is not optimized for microsecond-scale operations. 
We gave up one degree of freedom and designed {\leap} \cite{leap} in 2018 -- we optimized the OS for remote memory data path by identifying and removing non-critical operations while paging.

Even with the leanest data path, a reactive page fetching system must suffer microsecond-scale network latency on the critical path.
Leap introduced a remote memory prefetcher to proactively bring in the correct pages into a local cache to provide sub-microsecond latency (comparable to that of a local page access) on cache hits.


\paragraph{Providing Resilience.}
Infiniswap originally relied on local disks to tolerate remote failures, which resulted in slow failure recovery. 
Maintaining multiple in-memory replicas was not an option either as it effectively halved the total capacity. 
We started exploring erasure coding as a memory-efficient alternative.
Specifically, we divided each page into $k$ splits to generate $r$ encoded parity splits and spread the $(k + r)$ splits to $(k + r)$ failure domains -- any $k$ out of $(k + r)$ splits would then suffice to decode the original data. 
However, erasure coding was traditionally applied to large objects \cite{ec-cache}.
By 2019/20, we built \hydra \cite{hydra} whose carefully designed data path could perform online erasure coding within a single-digit microsecond tail latency.
{\hydra} also introduced CodingSets, a new data placement scheme that balanced availability and load balancing, while reducing the probability of data loss by an order of magnitude even under large correlated failures.

\paragraph{Multi-Tenancy Issues.}
We observed early on (circa 2017) that accessing remote memory over a shared network suffers from contention in the NIC and inside the network \cite{fairdma}. 
While our optimized data paths in Leap and Hydra could address some of the challenges inside the host, they did not extend to resource contentions in the RDMA NIC (RNIC). 
We designed \justitia \cite{justitia} in 2020 to improve the network bottleneck in RNICs by transparently monitoring the latency profiles of each application and providing performance isolation.
More recently, we have looked into improving Quality-of-Service (QoS) inside the network as well \cite{aequitas:sigcomm22}.

\paragraph{Expanding to Public Clouds.}
While Infiniswap and related projects were designed for cooperative private datacenters, memory disaggregation in public clouds faces additional concerns. 
In 2021, we finished designing \memtrade \cite{memtrade} to harvest all the idle memory within virtual machines (VMs) -- be it unallocated, or allocated to an application but infrequently utilized, and exposed them to a disaggregated memory marketplace.
\memtrade allows producer VMs to lease  their idle application memory to remote consumer VMs for a limited period of time while ensuring confidentiality and integrity. 
It employs a broker to match producers with consumers while satisfying performance constraints.

\begin{figure}[!t]
	\centering
	\subfloat[][\textbf{Without CXL}]{
	\label{fig:cxl-less-bw}
		\includegraphics[width=0.7\columnwidth]{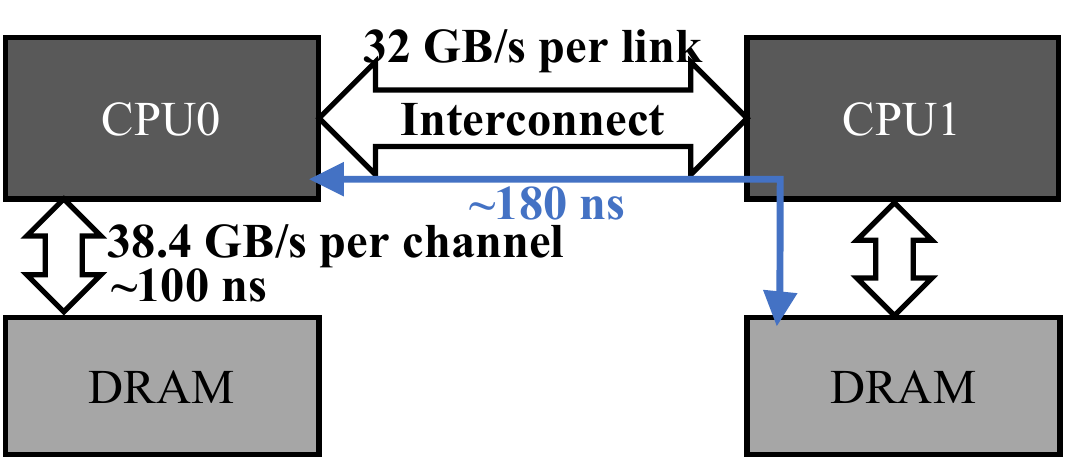}
	}

	\subfloat[][\textbf{With CXL on PCIe 5.0}]{
	\label{fig:cxl-bw}
		\includegraphics[width=0.7\columnwidth]{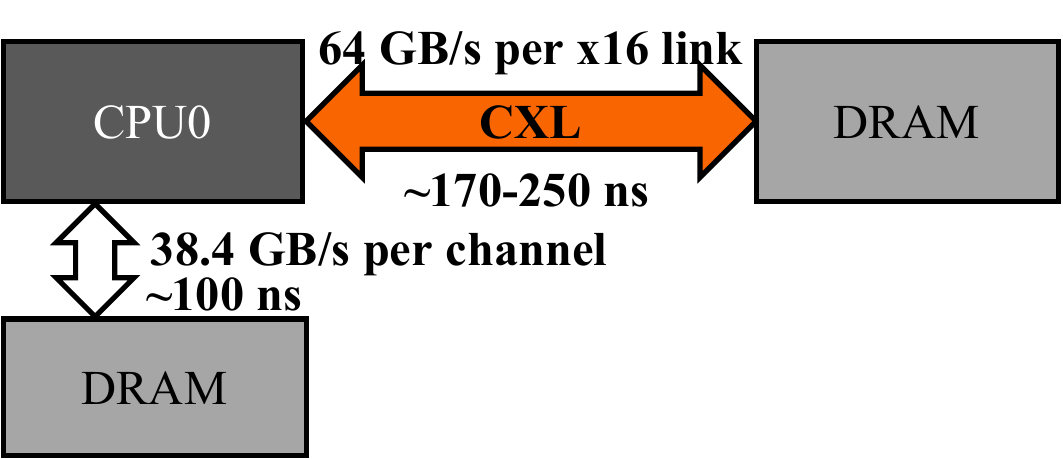}
	}
	\caption{A CXL system compared to a dual-socket server.}
	\label{fig:cxl-numa-alike}
	\vspace{-1em}
\end{figure}

\begin{figure*}[!t]
  \centering
  \includegraphics[width=1.3\columnwidth]{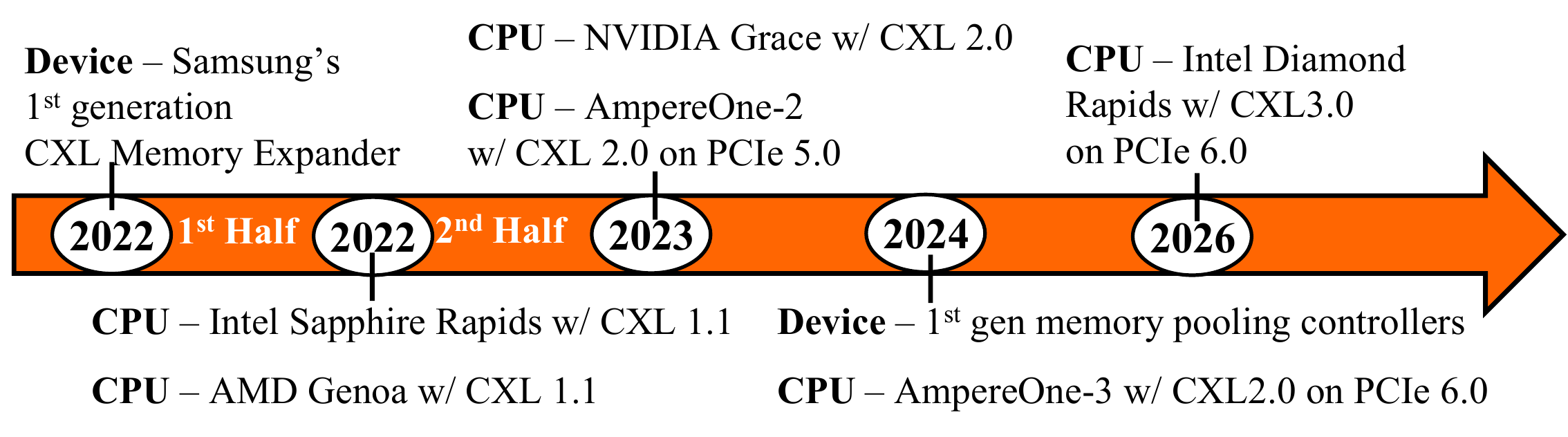}
  \caption{CXL roadmap paves the way for memory pooling and disaggregation in next-generation datacenter design.}
  \label{figure:roadmap}
  \vspace{-1em}
\end{figure*}

\paragraph{Detours Along the Way.}
Throughout this journey, we collaborated on side quests like designing a decentralized resource management algorithm using RDMA primitives \cite{dslr}, meta-data management inside the network using programmable switches \cite{netlock:sigcomm20}, fine-grained compute disaggregation \cite{kayak:nsdi21} \etc
Some of our forays into designing hardware support were nipped in the bud, often because we could not find the right partners. 
In hindsight, perhaps we were fortunate given how quickly the industry converged on CXL.

\paragraph{Summing it Up.}
Infiniswap along with all its extensions can provide near-memory performance for most memory-intensive applications even when 75\% and sometimes more of their working sets reside in remote memory in an application- and hardware-transparent manner, in the presence of failures, load imbalance, and multiple tenants.
After seven years, we declared victory on this chapter in 2022.

\section{Hardware Trend: Cache-Coherent\\ Interconnects}
\label{sec:cxl}
Although networking technologies like InfiniBand and Ethernet continue to improve, their latency remain considerably high for providing a cache-coherent memory address space across disaggregated memory devices.
CXL (Compute Express Link) \cite{cxl} is a new processor-to-peripheral/accelerator cache-coherent interconnect protocol that builds on and extends the existing PCIe protocol by allowing coherent communication between the connected devices.\footnote{Prior industry standards in this space such as CCIX \cite{ccix}, OpenCAPI \cite{opencapi}, Gen-Z \cite{genz} \etc have all come together under the banner of CXL consortium.
While there are some related research proposals (\eg, \cite{mind}), CXL is the de facto industry standard at the time of writing this paper.} 
It provides byte-addressable memory in the same physical address space and allows transparent memory allocation using standard memory allocation APIs. 
It also allows cache-line granularity access to the connected devices and underlying hardware maintains cache-coherency and consistency. 
With PCIe 5.0, CPU-to-CXL interconnect bandwidth is
similar to the cross-socket interconnects (Figure~\ref{fig:cxl-numa-alike}) on a dual-socket machine \cite{tioga-pass}. 
\cxlmem access latency is also similar to the NUMA access latency. 
CXL adds around 50-100 nanoseconds of extra latency over normal DRAM access.

\paragraph{CXL Roadmap.}
Today, CXL-enabled CPUs and memory devices support CXL 1.0/1.1 (Figure~\ref{figure:roadmap}) that enables a point-to-point link between CPUs and accelerator memory or between CPUs and memory extenders.
CXL 2.0 spec enables one-hop switching that allows multiple accelerators without (\emph{Type-1 device}) or with memory (\emph{Type-2 device}) to be configured to a single host and have their caches be coherent to the CPUs.
It also allows memory pooling across multiple hosts using memory expanding devices (\emph{Type-3 device}). 
A CXL switch has a fabric manager (it can be on-board or external) that is in charge of the device address-space management. 
Devices can be hot-plugged to the switch. 
A virtual CXL switch partitions the \cxlmem and isolate the resources between multiple hosts. 
It provides telemetry for load on each connected devices for load balancing and QoS management. 

\begin{figure}[!t]
	\includegraphics[width=\columnwidth]{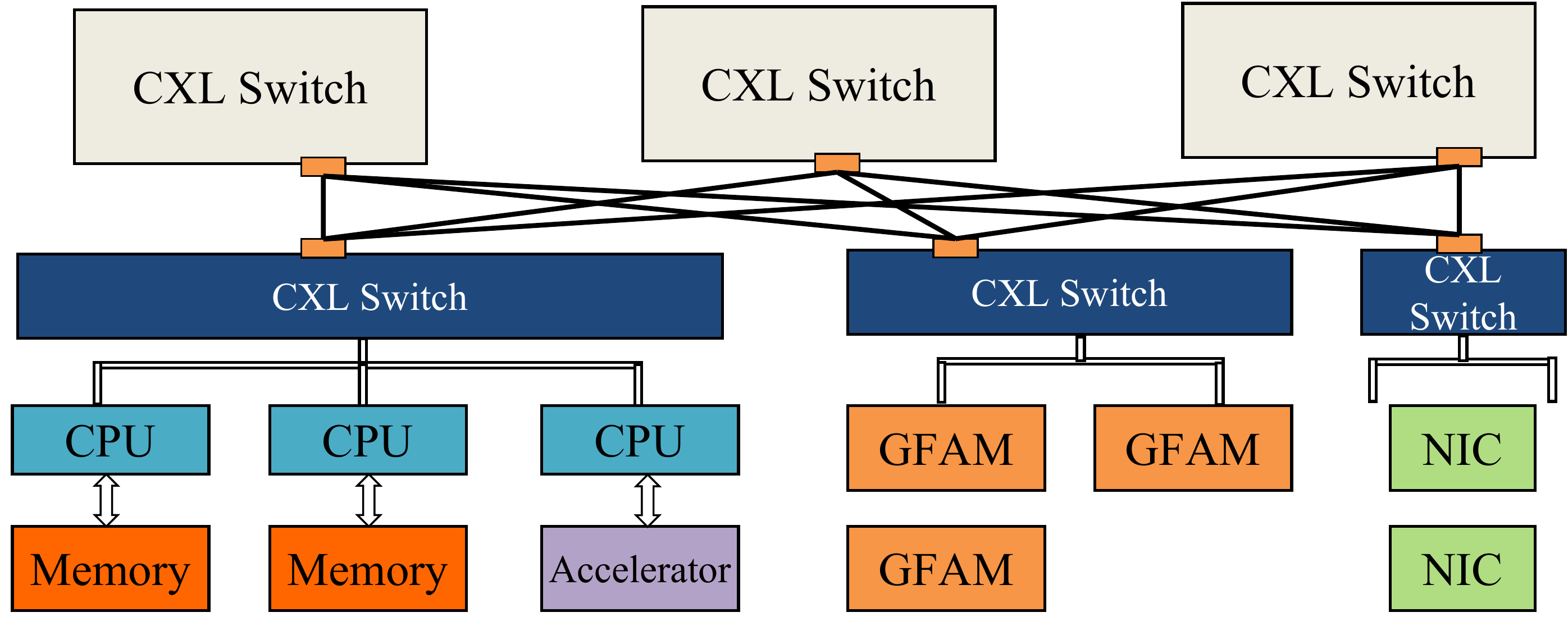}
	\caption{CXL 3.0 enables a rack-scale server design with complex networking and composable memory hierarchy.}
	\label{fig:cxl3}
	\vspace{-1.4em}
\end{figure}

CXL 3.0 adds multi-hop hierarchical switching -- one can have any complex types of network through cascading and fan-out. 
This expands the number of connected devices and the complexity of the fabric to include non-tree topologies, like Spine/Leaf, mesh- and ring-based architectures. 
CXL 3.0 supports PCIe 6.0 (64 GT/s \ie, up to 256 GB/s of throughput for a x16 duplex link) and expand the horizon of very complex and composable rack-scale server design with varied memory technologies (Figure~\ref{fig:cxl3}).
A new Port-Based Routing (PBR) feature provides a scalable addressing mechanism that supports up to 4,096 nodes. 
Each node can be any of the existing three types of devices or the new Global Fabric Attached Memory (GFAM) device  that supports different types of memory (\ie, Persistent Memory, Flash, DRAM, other future memory types, \etc) together in a single device. 
Besides memory pooling, CXL 3.0 enables memory sharing across multiple hosts on multiple end devices. 
Connected devices (\ie, accelerators, memory expanders, NICs, \etc) can do peer-to-peer communicate bypassing the host CPUs. 

In essence, CXL 3.0 enables large networks of memory devices. 
This will proliferate software-hardware co-designed memory disaggregation solutions that not only simplify and better implement previous-generation disaggregation solutions (\eg, Infiniswap) but also open up new possibilities. 

\section{Disaggregation Over Intra-Server CXL}
\label{sec:tpp}

With the emergence of new hardware technologies comes the opportunity to rethink and revisit past design decisions, and CXL is no different.
Earlier software solutions for memory disaggregation over RDMA are not optimized enough in CXL-based because of its much lower latency bound, especially for intra-server CXL (CXL 1.0/1.1) with 100s of nanoseconds latency.
%
Recent works in leveraging CXL 1.0/1.1 within a server have focused on (tiered) memory pooling \cite{tpp,pond-azure} because a significant portion of datacenter application working sets can be offloaded to a slower-tier memory without hampering performance \cite{googledisagg, tpp, memtrade}.
We have recently worked on two fundamental challenges in this context.

\paragraph{Memory Usage Characterization.}
Datacenter applications have diverse memory access latency and bandwidth requirements.
Sensitivity toward different memory page types can also vary across applications.
Understanding and characterizing such behaviors is critical to designing heterogeneous tiered-memory systems.
%
{\chameleon} \cite{tpp} is a lightweight user-space memory access behavior characterization tool that can readily be deployed in production without disrupting running application(s) or modifying the OS.
It utilizes the Precise Event-Based Sampling (PEBS) mechanism of modern CPU's Performance Monitoring Unit (PMU) to collect hardware-level performance events related to memory accesses.
It then generates a heat-map of memory usage for different page types and provides insights into an application's expected performance with multiple temperature tiers.

\paragraph{Memory Management.}
Given applications' page characterizations, \tpp \cite{tpp} provides an OS-level transparent page placement mechanism, to efficiently place pages in a tiered-memory system.
\tpp has three components: \textbf{(a)} a lightweight reclamation mechanism to demote colder pages to the slow tier; \textbf{(b)} decoupling the allocation and reclamation logic for multi-NUMA systems to maintain a headroom of free pages on the fast tier; and \textbf{(c)} a reactive page promotion mechanism that efficiently identifies hot pages trapped in the slow memory tier and promote them to the fast memory tier to improve performance.
It also introduces support for page type-aware allocation across the memory tiers.



\section{CXL-Disaggregated Memory at Rack-Scale and Beyond: Open Challenges}

Although higher than intra-server CXL latency, rack-scale CXL systems with a CXL switch (CXL 2.0) will experience much lower latency than RDMA-based memory disaggregation.
With a handful of hops in CXL 3.0 setups, latency will eventually reach a couple microseconds similar to that found in today's RDMA-based disaggregated memory systems.
For next-generation memory disaggregation systems that operate between these two extremes, \ie, rack-scale and a little beyond, many open challenges exist.
We may even have to revisit some of our past design decisions (\S\ref{sec:overview}).
Here we present a non-exhaustive list of challenges informed by our experience.

\subsection{Abstractions}
\paragraph{Memory Access Granularity.} 
CXL enables cache-line granular memory access over the connected devices, whereas existing OS VMM modules are designed for page-granular (usually, 4KB or higher) memory access.
Throughout their lifetimes, applications often write a small part of each page; typically only 1-8 cache-lines out of 64 \cite{kona}.
Page-granular access causes large dirty data amplification and bandwidth overuse.
In contrast, fine-grained memory access over a large memory pool causes high meta-data management overhead.
Based on an application's memory access patterns, remote memory abstractions should support transparent and dynamic adjustments to memory access granularity.

\paragraph{Memory-QoS Interface.}
Traditional solutions for memory page management focus on tracking (a subset of) pages and counting accesses to determine the heat of the page and then moving pages around. 
While this is enough to provide a two-level, hot-vs-cold QoS, it cannot capture the entire spectrum of page temperature. 
Potential solutions include assigning a QoS level to (1) an entire application; (2) individual data structures; (3) individual \texttt{mmap()} calls; or even (4) individual memory accesses. 
Each of these approaches have their pros and cons. 
At one extreme, assigning a QoS level to an entire application maybe simple, but it cannot capture time-varying page temperature of large, long-running applications.
At the other end, assigning QoS levels to individual memory accesses requires recompilation of all existing applications as well as cumbersome manual assignments, which can lead to erroneous QoS assignments.
A combination of aforementioned approaches may reduce  developer's overhead while providing sufficient flexibility to perform spatiotemporal memory QoS management.

\subsection{Management and Runtime}

\paragraph{Memory Address Space Management.}
From CXL 2.0 onward, devices can be hot-plugged to the CXL switches. 
Device-attached memory is mapped to the system's coherent address space and accessible to host using standard write-back semantics. 
Memory located on a CXL device can either be mapped as Host-managed Device Memory (HDM) or Private Device Memory (PDM).
To update the memory address space for connected devices to different host devices, a system reset is needed; traffic towards the device needs to stop to alter device address mapping during this reset period.
An alternate solution to avoid this system reset is to map the whole physical address space to each host when a CXL-device is added to the system.
The VMM or fabric manager in the CXL switch will be responsible to maintain isolation during address-space management.
How to split the whole address-space in to sizable memory blocks for the efficient physical-to-virtual address translation of a large memory network is an interesting challenge \cite{mind, netlock:sigcomm20}.

\paragraph{Unified Runtime for Compute Disaggregation.}
CXL Type-2 devices (accelerator with memory) maintains cache coherency with the CPU.
CPU and Type-2 devices can interchangeably use each other's memory and both get benefited. 
For example, applications that run on CPUs can benefit as they can now access very high bandwidth GPU memory.
Similarly, for GPU users, it is beneficial for capacity expansion even though the memory bandwidth to and from CPU memory will be lower.
In such a setup, remote memory abstractions should track the availability of compute cores and efficiently perform near-memory computation to improve the overall system throughput. 

Future datacenters will likely be equipped with numerous domain-specific compute resources/accelerators. 
In such a system, one can borrow the idle cores of one compute resource and perform extra computation to increase the overall system throughput.
A unified runtime to support malleable processes that can be immediately decomposed into smaller pieces and offloaded to any available compute nodes can improve both application and cluster throughput \cite{nu, kayak:nsdi21}.

\subsection{Allocation Policies}

\paragraph{Memory Allocation in Heterogenous NUMA Cluster.}
For better performance, hottest pages need to be on the fastest memory tier. 
However, due to memory capacity constraints across different tiers, it may not always be possible to utilize the fastest or performant memory tier.
Determining what fraction of memory is needed at a particular memory tier to maintain the desired performance of an application at different points of its life cycle is challenging. 
This is even more difficult when multiple applications coexist.
Efficient promotion or demotion of pages of different temperatures across memory tiers at rack scale is necessary.
One can consider augmenting \tpp by incorporating a lightweight but effective algorithm to select the migration target considering node distances from the CPU, load on CPU-memory bus, current load on different memory tiers, network state, and the QoS requirements of the migration-candidate pages.


\paragraph{Allocation Policy for Memory Bandwidth Expansion.}
For memory bandwidth-bound applications, CPU-to-DRAM bandwidth often becomes the bottleneck and increases the average memory access latency. 
CXL's additional memory bandwidth can help 
by spreading memory across the top-tier and remote nodes. 
Instead of only placing cold pages into \cxlmem, which has low bandwidth consumption, an ideal solution should place the right amount of bandwidth-heavy, latency-insensitive pages to \cxlmem. 
The methodology to identify the ideal fraction of such working sets may even require hardware support.

\paragraph{Memory Sharing and Consistency.}
CXL 3.0 allows memory sharing across multiple devices. 
Through an enhanced coherency semantics, multiple hosts can have a coherent copy of a shared segment, with back invalidation for synchronization.
Memory sharing improves application-level performance by reducing unnecessary data movement and improves memory utilization.
Sharing a large memory address space, however, results in significant overhead and complexity in the system that plagued classic distributed shared memory (DSM) proposals \cite{dsm-survey}.
Furthermore, sharing memory across multiple devices increases the security threat in the presence of any malicious application run on the same hardware space.
We believe that disaggregated memory systems should cautiously approach memory sharing and avoid it unless it is absolutely necessary for specific scenarios.

\subsection{Rack-Level Objectives}

\paragraph{Rack-Scale Memory Temperature.}
To obtain insights into an application's expected performance with multiple temperature tiers, it is necessary to understand the heat map of memory usage for that application. 
Existing hot page identification mechanisms (including \chameleon) are limited to a single host OS or user-space mechanism.
They either use access bit-based mechanism \cite{damon-tool, ipt-tool, idle-mem}, special CPU feature-based (\eg, Intel PEBS) tools  \cite{hemem, unimem, pebs-hybrid-memory}, or OS features \cite{autonuma, tpp} to determine the page temperature within a single server.
So far, there is no distributed mechanism to determine the cluster-wide relative page temperature. 
Combining the data of all the OS or user-space tools and coordinating between them to find rack-level hot pages is an important problem.
CXL fabric manager is perhaps the place where one can get a cluster-wide view of hardware counters for each CXL device's load, hit, and access-related information. 
One can envision extending \chameleon for rack-scale environments to provide observability into each application's per-device memory temperature.

\paragraph{Hardware-Software Co-Design for a Better Ecosystem.}
Hardware features can further enhance performance of disaggregation systems in rack-scale setups. 
A memory-side cache and its associated prefetcher on the CXL ASIC or switch might help reduce the effective latency of \cxlmem.
Hardware support for data movement between memory tiers can help reduce page migration overheads in an aggressively provisioned system with very small amount of local memory and high amount of \cxlmem. 
Additionally, the fabric manager of a CXL switch should implement policies like fair queuing, congestion control, load balancing \etc for better network management.
Incorporating \leap's prefetcher and \hydra's erasure-coded resilience ideas into CXL switch designs can enhance system-wide performance.


\paragraph{Energy- and Carbon-Aware Memory Disaggregation.}
Datacenters represent a large and growing source of energy consumption and carbon emissions \cite{treehouse:hotcarbon22}. 
Some estimates place datacenters to be responsible for 1-2\% of aggregate worldwide electricity consumption \cite{datacenter-electricity, energy-hogs}.
To reduce the TCO and carbon footprint, and enhance hardware life expectancy, datacenter rack maintain a physical energy budget or power cap.
Rack-scale memory allocation, demotion, and promotion policies can be augmented by incorporating energy-awareness in their decision-making process. 
In general, we can introduce energy-awareness in the software stack that manage compute, memory, and network resources in a disaggregated cluster. 

\section{Conclusion}
\label{sec:outro}
We started the Infiniswap project in 2016 with the conviction that memory disaggregation is inevitable, armed only with a few data points that hinted it might be within reach.
As we conclude this paper in 2023, we have successfully built a comprehensive software-based disaggregated memory solution over ultra-fast RDMA networks that can provide a seamless experience for most memory-intensive applications.
With diverse cache-coherent interconnects finally converging under the CXL banner, the entire industry (and ourselves) are at the cusp of taking a leap toward next-generation software-hardware co-designed disaggregated systems. 
Join us.
Memory disaggregation is here to stay.

\section*{Acknowledgements}
Juncheng Gu, Youngmoon Lee, and Yiwen Zhang co-led different aspects of the Infiniswap project alongside the authors. 
We thank Yiwen Zhang for his feedback on this paper.
Special thanks to our many collaborators, contributors, users, and cloud resource providers (namely, CloudLab, Chameleon Cloud, and UM ConFlux) for making Infiniswap successful.
Our expeditions into next-generation memory disaggregation solutions have greatly benefited from our collaborations with Meta. 
Our research was supported in part by National Science Foundation grants (CCF-1629397, CNS-1845853, and CNS-2104243) and generous gifts from VMware and Meta.

\label{lastpage}

\bibliographystyle{abbrv}
\bibliography{memory_disaggregation}
\end{document}